\documentclass[preprint2,10.5pt]{aastex}
\begin{document}
\title{\large{\rm{THE PULSATION MODE OF THE CEPHEID POLARIS}}}
\author{D. G. Turner$^1$, V. V. Kovtyukh$^{2,3}$, I. A. Usenko$^{2,3}$, N. I. Gorlova$^4$}
\affil{$^1${\rm \small Saint Mary's University, Halifax, Nova Scotia, Canada.}}
\affil{$^2${\rm \small Astronomical Observatory, Odessa National University, Odessa, Ukraine.}}
\affil{$^3${\rm \small Isaac Newton Institute of Chile, Odessa Branch, Odessa, Ukraine.}}
\affil{$^4${\rm \small Institute of Astronomy, Leuven, Belgium.}}
\email{\rm turner@ap.smu.ca}

\begin{abstract}
A previously-derived photometric parallax of $10.10\pm0.20$ mas, $d = 99\pm2$ pc, is confirmed for Polaris by a spectroscopic parallax derived using line ratios in high dispersion spectra for the Cepheid. The resulting estimates for the mean luminosity of $\langle M_V \rangle = -3.07 \pm 0.01$ s.e., average effective temperature of $\langle {\rm T}_{\rm eff}\rangle = 6025 \pm 1$ K s.e., and intrinsic color of $(\langle B \rangle - \langle V \rangle)_0 = +0.56 \pm 0.01$ s.e., which match values obtained previously from the photometric parallax for a space reddening of $E_{B-V}=0.02\pm0.01$, are consistent with fundamental mode pulsation for Polaris and a first crossing of the instability strip, as also argued by its rapid rate of period increase. The systematically smaller Hipparcos parallax for Polaris appears discrepant by comparison.
\end{abstract}

\keywords{stars: variables: Cepheids---stars: distances---stars: individual (Polaris).}

\section{{\rm \footnotesize INTRODUCTION}}

Our current knowledge of the intrinsic properties of classical Cepheid variables relies heavily on the observational parameters derived for them. In the case of the nearest Cepheid, Polaris \citep[$\alpha$ UMi, $P=3.969$ days, see][]{te05}, the distance and reddening are of paramount importance for understanding its pulsation mode and evolutionary status. A well-defined space reddening of $E_{B-V}=0.02\pm0.01$ is implied by studies of its optical companion and other stars lying in the immediate vicinity of the Cepheid \citep{tu77,gf78,tu06,tu09}, but the distance remains a point of contention.

A photometric parallax can be inferred for Polaris using its $18\arcsec$-distant F3~V companion \citep{tu77}, which is recognized to be physically associated with the Cepheid on the basis of common proper motions and radial velocities \citep{ka96}. The distance derived from zero-age main-sequence (ZAMS) fitting for Polaris B is $101 \pm3$ pc \citep{tu06}, or 109.5 pc from a spectroscopic investigation \citep{us08}. An investigation of spatially-adjacent A, F, and G-type dwarfs within $3\degr$ of Polaris observed by Hipparcos \citep{esa97} reveals two distinct groups lying along the line of sight \citep{tu04}, of which only the closer contains stars of comparable proper motion and radial velocity to the Cepheid. The stars also concentrate spatially towards Polaris, and appear to constitute the remains of a sparse cluster in the final stages of dissolution into the Galactic disk. The implied distance from ZAMS fitting is $99\pm2$ pc \citep{tu06,tu09}, and the corresponding photometric parallax is $10.10\pm0.20$ mas.

With the above reddening, the photometric distance implies $\langle M_V\rangle=-3.07\pm0.04$ \citep{tu06}, consistent with fundamental mode pulsation for a 3.969-day Cepheid, along with a location near the center of the instability strip \citep*{te05,te06,tu09}. A potential conflict with the Cepheid's small light amplitude, more typical of stars on the hot and cool edges of the strip, was resolved by \citet{tu06,tu09} using the observation that Polaris appears to be in the first crossing of the instability strip. The Cepheid's implied location near strip center, despite an extremely low amplitude, can be attributed to a narrow and blueward-skewed instability strip for first crossers, in which surface convection damps pulsation at significantly warmer effective temperatures than for other crossings \citep[see][]{al99}. The redwards evolution of Polaris towards the cool edge of the instability strip for first crossers implied by its steady period increase \citep{te05,tu09} is also consistent with its apparently decreasing light amplitude \citep{tu09}.

Results for the trigonometric parallax of Polaris differ from those implied by its photometric parallax. Refractor parallaxes summarized by \citet*{va95} yield a parallax of $4.0\pm3.3$ mas for Polaris, but that does not appear to account for a magnitude dependence in the original Allegheny parallaxes, attributed to use of a rotating sector to diminish the flux from bright stars \citep{ha78}. If the magnitude-dependent correction found by \citet{hl83} is applied, the older parallaxes summarized by \citet{je52,je63}, calibrated relative to Allegheny parallaxes \citep[e.g.,][]{wa56}, yield a best value of $11\pm4$ mas for the trigonometric parallax of Polaris, in agreement with the photometric result.

In contrast, the Hipparcos parallax of $7.56\pm0.48$ mas \citep{esa97}, or $7.54\pm0.11$ mas from the new reduction \citep{vl07}, implies a distance of $133\pm2$ pc to Polaris. At that distance with the reddening cited previously, the luminosity of the Cepheid is $\langle M_V\rangle=-3.62\pm0.05$, which implies overtone pulsation, consistent with its sinusoidal light curve and small amplitude. But the intrinsic color of $(\langle B \rangle - \langle V \rangle)_0 = +0.56 \pm 0.01$ still leaves Polaris well inside the hot edge of the instability strip, and its rapid rate of period increase implies either a first or higher than third crossing of the instability strip, with an expected increasing light amplitude. That conflicts with the results for other small amplitude Cepheids as well as the long-term decreasing light amplitude observed prior to the 1965 ``glitch'' \citep{tu09}. In addition, the A, F, and G-type dwarfs in the vicinity of Polaris lying at distances comparable to that inferred from the Hipparcos parallax do not share the proper motion or radial velocity of the Cepheid \citep{tu04}, producing further inconsistencies.

Such contradictions were overlooked when \citet{fc97} and \citet{vl08} used Polaris as an overtone pulsator to calibrate the Cepheid period-luminosity relation using Hipparcos parallaxes. The Hipparcos parallax was also adopted by \citet{we00} and \citet{ev08} in their analyses of the orbit of the F6~V radial velocity companion, Polaris Ab, as well as by \citet{no00} with the star's measured angular diameter to estimate its radius. The process can be inverted for the angular diameter, however, and in combination with the well-established period-radius relation for classical Cepheids \citep{tb02,te10} yields distance estimates of $93\pm2$ pc ($\theta_{\rm LD}$) and $97\pm2$ pc ($\theta_{\rm UD}$) for fundamental mode pulsation, and $122\pm3$ pc ($\theta_{\rm LD}$) and $128\pm3$ pc ($\theta_{\rm UD}$) for overtone pulsation, where $\theta_{\rm LD}$ and $\theta_{\rm UD}$ apply to the limb-darkened and uniform disk solutions, respectively. Agreement with the results from the parallax solutions is less than ideal. Orbital radial velocity residuals \citep{le08} and tests of possible parameters for close orbital companions to Polaris \citep{tu09} also suggest the possibility of an extra star in the Polaris A subsystem, so an independent test of the distance, luminosity, and pulsation mode of the Cepheid would be useful.

Spectroscopic parallaxes can be derived for Cepheids using line ratios, the basis of temperature and luminosity ($\log g$) discrimination in Morgan and Keenan (MK) spectral classification \citep{gc09}. Kovtyukh and his collaborators \citep{ko07,ko10,ko12a,ko12b} have taken the technique to its natural limits by using high dispersion spectra and all possible line ratios, in conjunction with calibrations on T$_{\rm eff}$ and $M_V$, to establish the luminosities and effective temperatures of yellow supergiants and Cepheids with very high precision. Deatils of the calibration philosophy using line ratios involving primarily iron peak elements in the spectra of supergiants of well established luminosity are provided by \citet{ko09}. Typically the precision averages $\pm0^{\rm m}.26$ in absolute visual magnitude ($M_V$) for a single line ratio in Cepheids, but with so many line ratios available per spectrum, of order 40--70 \citep{ko12b}, the precision reached per spectrum reaches a few hundredths of a magnitude.

The method, although calibrated using yellow supergiants, was established with the aim of studying Cepheid variables, and was used recently by \citet{ko12b} to examine the pulsation modes of three s-Cepheids, with fairly robust results: V1334 Cyg (first overtone), V440 Per (fundamental mode), and V636 Cas (fundamental mode). It is ideal for learning more about Polaris, another s-Cepheid of extremely small amplitude and an object that is bright and readily accessible from northern hemisphere sites. The present study addresses the ongoing problem of the distance and pulsation mode of Polaris using its spectroscopic parallax.

\begin{deluxetable}{@{\extracolsep{-2.5mm}}cccccccccc}
\tabletypesize{\small}
\tablewidth{0pt}
\tablecaption{Spectroscopic Results for Polaris \label{tab1}}
\tablehead{
\colhead{JD(obs)} &\colhead{Phase} &\colhead{$T_{\rm eff}$} &\colhead{s.d.} &\colhead{n}  &\colhead{s.e.} &\colhead{$M_V$} &\colhead{s.d.} &\colhead{n} &\colhead{s.e.} \\
& &\colhead{(K)} &\colhead{(K)} &  &\colhead{(K)} & & & & } 
\startdata
2452861.5600 &0.6293 &6033 &59 &61 &7.5 &--3.03 &0.20 &55 &0.03 \\
2452867.5620 &0.1402 &6009 &35 &57 &4.6 &--2.98 &0.17 &52 &0.02 \\
2452869.5700 &0.6456 &5996 &61 &62 &7.7 &--3.02 &0.24 &44 &0.04 \\
2453072.1650 &0.6452 &6015 &74 &56 &9.9 &--3.05 &0.27 &67 &0.03 \\
2453073.6220 &0.0120 &6013 &88 &56 &11.8 &--3.06 &0.21 &66 &0.03 \\
2453162.1910 &0.3075 &6050 &61 &57 &8.1 &--3.05 &0.25 &67 &0.03 \\
2453689.6470 &0.0829 &6050 &49 &74 &5.6 &--3.10 &0.33 &62 &0.04 \\
2453690.1090 &0.1992 &6034 &42 &74 &4.9 &--3.06 &0.23 &62 &0.03 \\
2453693.1240 &0.9582 &6044 &61 &72 &7.2 &--3.01 &0.34 &68 &0.04 \\
2453980.5890 &0.3201 &6018 &41 &73 &4.8 &--3.04 &0.16 &62 &0.02 \\
2454073.5890 &0.7303 &6051 &46 &55 &6.2 &--3.04 &0.12 &53 &0.02 \\
2454077.6510 &0.7528 &6093 &125 &5 &55.9 &--3.17 &0.15 &8 &0.05 \\
2454169.6380 &0.9080 &6081 &97 &6 &39.4 &--3.03 &0.13 &10 &0.04 \\
2454225.2280 &0.9012 &6069 &27 &57 &3.6 &--3.02 &0.15 &60 &0.02 \\
2454345.5510 &0.1889 &6073 &35 &62 &4.4 &--3.07 &0.16 &61 &0.02 \\
2454426.0180 &0.4440 &5993 &47 &71 &5.6 &--3.11 &0.23 &65 &0.03 \\
2454934.5880 &0.4598 &6005 &52 &63 &6.5 &--3.01 &0.22 &60 &0.03 \\
2455005.3720 &0.2772 &5982 &45 &74 &5.2 &--3.14 &0.20 &65 &0.03 \\
2455324.6730 &0.6493 &5998 &47 &64 &5.9 &--3.11 &0.15 &70 &0.02 \\
2455328.5976 &0.6372 &5990 &41 &66 &5.1 &--3.09 &0.19 &64 &0.02 \\
2455708.3364 &0.2212 &5996 &52 &61 &6.6 &--3.06 &0.21 &69 &0.03 \\
2455816.5457 &0.4583 &5959 &67 &75 &7.7 &--3.10 &0.20 &63 &0.03 \\
2455901.6061 &0.8686 &6017 &79 &74 &9.2 &--3.18 &0.17 &62 &0.02 \\
\enddata
\end{deluxetable}

\begin{figure}[!t]
\epsscale{0.9}
\plotone{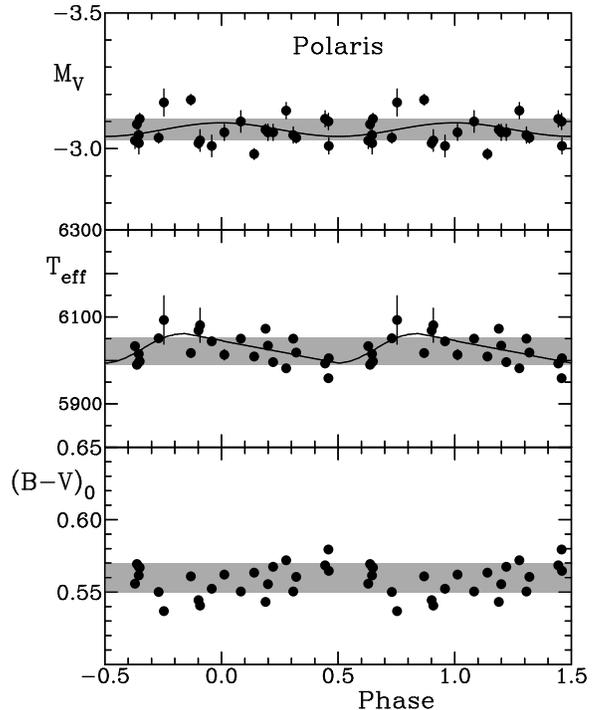}
\caption{\small{The luminosity (top), T$_{\rm eff}$ (middle), and predicted color variations (bottom) for Polaris from the spectra analyzed here, with uncertainties in the data indicated. The gray bands represent the range of values predicted for Polaris from its photometric parallax and space reddening \citep{tu06}, with superposed curves representing the light variations typical of the era of observation and a Fourier fit to the T$_{\rm eff}$ data.}}
\label{fig1}
\end{figure}

 \section{{\rm \footnotesize OBSERVATIONS AND DATA REDUCTIONS}}

Observations of Polaris were obtained using the 6-m Large Azimuth Telescope (BTA) of the Special Astrophysical Observatory of the Russian Academy of Sciences equipped with the Nasmyth Echelle Spectrometer \citep[NES,][]{pa09}, which has a resolving power of ${\rm R}\approx60\,000$ within the wavelengh range 4380--6690\,\AA. The signal-to-noise ratio at the continuum level in each of the 20--27 spectral orders exceeds 70. A thorium-argon lamp was used for wavelength calibration, and the data reduction was carried out using the MIDAS software ECHELLE modified for extraction of echelle spectra obtained with an image slicer \citep{yu05}. The spectra were extracted from the CCD images in the usual fashion: bias subtraction, flat-fielding, cosmic ray removal, and wavelength calibration. 

Spectra were also obtained with the fiber echelle-type spectrograph HERMES, mounted on the 1.2-m Belgian telescope on La Palma. A high-resolution configuration with  R = 85\,000 for the wavelength region 3800--9000 \AA\ was used. The spectra were reduced using a Python-based pipeline that includes order extraction, wavelength calibration with Th-Ne-Ar arcs, flat field division, cosmic-ray clipping, and order merging. Further details of the spectrograph and pipeline are given by \citet{ra11}. The dates of observation are indicated in Table~\ref{tab1}.

\section{{\rm \footnotesize RESULTS}}

The results of the analysis are summarized in Table~\ref{tab1} and plotted as a function of pulsation phase in Fig.~\ref{fig1}, where the phases were calculated with the regular period increase removed \citep{te05,tu09}. The plotted gray bands are {\it not} least squares solutions to the data. They represent the values with their cited uncertainties determined for Polaris by \citet{tu06} from the photometric parallax in conjunction with the reddening inferred from its companion. The observations themselves are fairly randomly distributed in phase around the Cepheid's cycle, and yield weighted mean values of $\langle M_V \rangle = -3.07 \pm 0.01$ s.e., and $\langle {\rm T}_{\rm eff}\rangle = 6025 \pm 1$ K s.e., identical to predictions \citep{tu06}. They confirm the previous conclusion that Polaris is a fundamental mode pulsator. If it were pulsating in an overtone mode at the distance implied by its Hipparcos parallax, the observational data would yield significantly greater luminosities closer to $M_V\simeq -3.6$.

Curves in Fig.~\ref{fig1} represent the expected sinusoidal variations in visual light and a Fourier fit to the effective temperature estimates. The data are precise enough to track pulsational changes in the latter. The star reaches highest effective temperature two tenths of a cycle prior to maximum light near minimum radius, and lowest effective temperature seven tenths of a cycle later near maximum radius. The temperature variations are skewed, unlike the more sinusoidal light variations, with a variation of $\sim75$ K in temperature over the course of a cycle. The color variations predicted by the estimates of effective temperature \citep[see][]{tb02,te10} mimic the temperature changes in displaying skewness, and yield an unweighted mean value of $(\langle B \rangle - \langle V \rangle)_0 = +0.56 \pm 0.01$ s.e. for Polaris. By comparison, the changes in absolute magnitude appear to be relatively small and sinusoidal, as observed for the visual light variations. Although the predicted and actual changes are similar, scatter in the individual estimates hinders more definitive conclusions. The trends are otherwise as expected for a fundamental mode pulsator.

The abundance patterns in Polaris --- [C/H] = --0.17, [N/H] = +0.42, [O/H] = --0.00, and [Na/H] = +0.09 \citep{us05} --- are those of a star displaying the products of core CNO processing at its surface. Sometimes that is taken to be a signature that a Cepheid has passed through the red supergiant stage, where deep envelope convective dredge-up is thought to bring core-processed material to the stellar surface \citep{mo99a,mo99b}. But some stellar evolutionary models do not involve a dredge-up stage for red supergiants \citep[e.g.,][]{bo00}, and \citet{ma01} has noted that most B-type stars in late main-sequence hydrogen burning stages prior to evolution towards the Cepheid instability strip already display CNO-processed material in their atmospheres. Rapid rotation during main-sequence stages in conjunction with meridional mixing of core material to the stellar surface is sufficient to explain the abundance patterns in Cepheids \citep[see][]{tb04}, and that may be the case for Polaris, i.e., its progenitor was a rapid rotator as a B-dwarf. The atmospheric abundances of Polaris A and B \citep{us05,us08} are otherwise indicative of slightly metal-rich stars with [Fe/H] = +0.07.

\section{{\rm \footnotesize POLARIS AS A FIRST-CROSSING CEPHEID}}

A potential problem with the spectroscopic parallax concerns the conclusion reached previously that Polaris is in the first crossing of the instability strip, as inferred from its observed rate of period increase of $\sim 4.5$ s yr$^{-1}$ \citep{te05,tu09}. \citet{ne12} argue that the value appears smaller than what is predicted by stellar evolutionary models \citep{te06}, and propose a higher strip crossing mode for Polaris that includes a component of the Cepheid's period increase arising from mass loss.

There is an alternate solution. In their comparison of observed rates of period change in Cepheids with those predicted from stellar evolutionary models, \citet{te06} employed a semi-empirical approach \citep{tu96,tb02} in which Cepheid radii varied as $P^{0.75}$ and masses as $P^{0.5}$, where {\it P} is the pulsation period. That yielded an equation for rate of period change of:
\begin{displaymath}
\frac{\dot{P}}{P} = \frac{6}{7}\frac{\dot{L}}{L} - \frac{24}{7}\frac{\dot{T}}{T}  
\end{displaymath}
where the quantities on the right hand side of the equation are taken from evolutionary models for stars crossing the Cepheid instability strip.

\begin{figure}[!t]
\epsscale{0.9}
\plotone{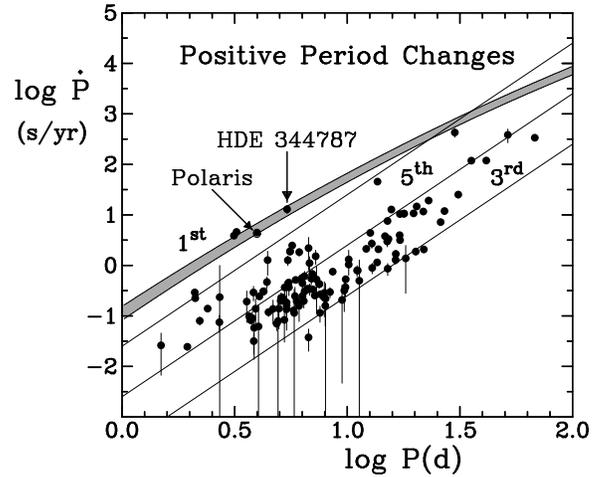}
\caption{\small{Observed and predicted rates of period increase for Cepheids in the first, third, and putative fifth crossings of the instability strip, i.e., Cepheids exhibiting period increases. Lines separate what appear to be stars in third and fifth crossings, and a gray band represents predictions for first-crossing Cepheids according to stellar evolutionary models (see text). The small amplitude Cepheids Polaris and HDE~344787 display period increases expected for a first crossing.}}
\label{fig2}
\end{figure}

More recent studies of Cepheids belonging to open clusters indicate that the $P^{0.5}$ dependence for Cepheid masses is too steep, and more likely varies as $P^{0.4}$ \citep{tu12}. With that adjustment the predicted rates of period change are modified to:
\begin{displaymath}
\frac{\dot{P}}{P} = \frac{5}{8}\frac{\dot{L}}{L} - \frac{5}{2}\frac{\dot{T}}{T} 
\end{displaymath}
The effect is to reduce the predicted rates of period change for Cepheids in the first crossing of the instability strip, as depicted in Fig.~\ref{fig2}. The observed rate of period increase for Polaris is now fully consistent with a first crossing of the instability strip, without the need to postulate mass loss or overtone pulsation for a different strip crossing. The observed rate falls so close to the minimum predicted rate, in fact, that mass loss in the Cepheid must be almost negligible.

\section{{\rm \footnotesize SUMMARY}}

An independent test of the phase-dependent variations in luminosity and effective temperature of Polaris is made using line ratios in high dispersion spectra for the star, calibrated using stars of similar metallicity to the Cepheid. The observational results are relevant to the star's distance and pulsation mode. The derived absolute magnitudes $M_V$ and effective temperatures T$_{\rm eff}$ coincide exactly with similar parameters predicted from the star's photometric parallax \citep{tu06}. The spectroscopic and photometric parallaxes both imply a distance to Polaris of $99\pm2$ pc. The associated Hipparcos parallax for Polaris \citep{esa97,vl07} appears to be discrepant by comparison.

The results are consistent with fundamental mode pulsation in Polaris, as well as with a first crossing of the instability strip. A correction of previously-published predictions for first-crossing Cepheids \citep{te06} to account for a more correct period-mass relation for Cepheids brings the observed rate of period increase in Polaris into good agreement with predictions from stellar evolutionary models.

\subsection*{{\rm \scriptsize ACKNOWLEDGEMENTS}}
\scriptsize{Some of the spectra were collected with the Mercator Telescope, operated on the island of La Palma by the Flemish Community, at the Spanish Observatorio del Roque de los Muchachos of the Instituto de Astrofisica de Canarias. The observatory staff and Drs.~V.~G. Klochkova and M.~V. Yushkin are acknowledged for their help with the spectral material. Much of the information about the supergiants used in the calibration was gathered with the help of SIMBAD. We are grateful to the referee for useful suggestions on the presentation of the paper.}

\end{document}